\documentclass[%
aip,
amsmath,amssymb,
reprint,%
]{revtex4-1}

\usepackage{graphicx}
\usepackage{dcolumn}
\usepackage{bm}
\usepackage[colorlinks, linkcolor=red, anchorcolor=blue,urlcolor=blue, citecolor=blue]{hyperref}
\usepackage{color}
\usepackage{mathrsfs}
\usepackage{amsfonts}
\usepackage{amssymb}
\usepackage{amsmath}
\usepackage[utf8]{inputenc}
\usepackage[T1]{fontenc}
\usepackage{mathptmx}

\newcommand{\beq}{\begin{equation}}
\newcommand{\eeq}{\end{equation}}
\def\bse{ \begin{subequations} }
	\def\ese{ \end{subequations} }
\def\beqa{ \begin{eqnarray} }
\def\eeqa{ \end{eqnarray} }

\begin{document}
	
	\title{Shortcuts to adiabaticity for an interacting Bose-Einstein condensate via exact solutions of the generalized Ermakov equation}

	\author{Tang-You Huang}
	\affiliation{International Center of Quantum Artificial Intelligence for Science and Technology (QuArtist) \\ and Department of Physics, Shanghai University, 200444 Shanghai, China}

	\author{Boris A. Malomed}
	\affiliation{Department of Physical Electronics, School of Electrical Engineering,
		Faculty of Engineering, Tel Aviv University, P.O.B. 39040, Ramat Aviv, Tel Aviv, Israel}
	\affiliation{Center for Light-Matter Interaction, Tel Aviv University, P.O.B. 39040, Ramat Aviv, Tel Aviv, Israel}
	
	\author{Xi Chen}
	\email{xchen@shu.edu.cn}
	\affiliation{International Center of Quantum Artificial Intelligence for Science and Technology (QuArtist) \\ and Department of Physics, Shanghai University, 200444 Shanghai, China}
	\affiliation{Department of Physical Chemistry, University of the Basque Country UPV/EHU, Apartado 644, 48080 Bilbao, Spain}
	
\begin{abstract}
Shortcuts to adiabatic expansion of the effectively one-dimensional
Bose-Einstein condensate (BEC) loaded in the harmonic-oscillator (HO) trap
is investigated by combining techniques of the variational approximation and
inverse engineering. Piecewise-constant (discontinuous) intermediate trap
frequencies, similar to the known bang-bang forms in the optimal-control
theory, are derived from an exact solution of a generalized Ermakov
equation.
Control schemes considered in the paper include imaginary trap
frequencies at short time scales, i.e., the HO potential replaced by the quadratic repulsive
one. Taking into regard the BEC's intrinsic nonlinearity, results are
reported for the minimal transfer time, excitation energy (which measures
deviation from the effective adiabaticity), and stability for the
shortcut-to-adiabaticity protocols. These results are not only useful for
the realization of fast frictionless cooling, but also help to address
fundamental problems of the quantum speed limit and thermodynamics.
\end{abstract}

\date{\today }
\maketitle

\author{Tang-You Huang}
\affiliation{International Center of Quantum Artificial Intelligence for Science and Technology (QuArtist) \\ and Department of Physics, Shanghai University, 200444 Shanghai, China}

\author{Boris A. Malomed}
\affiliation{Department of Physical Electronics, School of Electrical Engineering,
	Faculty of Engineering, Tel Aviv University, P.O.B. 39040, Ramat Aviv, Tel Aviv, Israel}
\affiliation{Center for Light-Matter Interaction, Tel Aviv University, P.O.B. 39040, Ramat Aviv, Tel Aviv, Israel}

\author{Xi Chen}
\email{xchen@shu.edu.cn}
\affiliation{International Center of Quantum Artificial Intelligence for Science and Technology (QuArtist) \\ and Department of Physics, Shanghai University, 200444 Shanghai, China}
\affiliation{Department of Physical Chemistry, University of the Basque Country UPV/EHU, Apartado 644, 48080 Bilbao, Spain}

\begin{quotation}
"Shortcuts to adiabaticity" (STA) for efficient transformation of trapped
nonlinear-wave modes are important tools which help to improve quality of
the transformation, simultaneously optimizing its efficiency. In this work,
we focus on the shortcuts for expansion of effectively one-dimensional
Bose-Einstein condensates (BECs), described by the Gross-Pitaevskii equation
(GPE) including the cubic self-interaction of the wave function and the
harmonic-oscillator (HO) trapping potential with a time-dependent strength.
We design simple but fast STA protocols, using the method of inverse
engineering, realized by means of the variational approximation applied to
the GPE. A generalized Ermakov equation, including an additional term
induced by the self-interaction of the BEC, is thus derived (the classical
second-order ordinary differential equation of this type was derived and
solved by Russian mathematician Ermakov about 150 years ago). Results of the
analysis help us to elaborate schemes for the time modulation of the HO-trap
frequency, admitting fast frictionless cooling of the expanding BEC in the
weak-interaction regime. In particular, the schemes based on the
\textquotedblleft bang" and \textquotedblleft bang-bang" forms, which are
well known in the optimal-control theory, are exemplified. The minimal
transformation time, time-averaged energy of excitations generated around
the expanding state, and stability of the dynamical regimes with attractive
and repulsive self-interactions are analyzed for various STA protocols. In
addition to direct applications to the expansion (or compression) of BEC,
the results are relevant for studies of the quantum speed limit and
manifestations of the third principle of thermodynamics in quantum systems
in general.
\end{quotation}

\section{Introduction}

Precise control and manipulations of non-interacting and interacting
Bose-Einstein condensates (BECs) in trapping potentials has well-known
significance to applications ranging from quantum simulations, information
processing, and quantum-enhanced metrology to atom interferometry \cite%
{cohen2011advances}. A particularly relevant example is the transfer a
quantum system from the ground state of one potential into that of another,
through its evolution governed by a specifically designed Hamiltonian. To
this end, slow adiabatic processes \cite{Adiabaticcooling}, as well as fast
shortcuts through intermediate states \cite{anderson1994tunneling}, Fourier
transform \cite{couvert2008optimal}, optimal control \cite%
{Rabitzatom,RabitzBEC,salamon2009maximum}, and machine learning \cite%
{henson2018approaching} have been exploited for the realization of fast
frictionless cooling and transport of cold atoms, trapped ions, and BEC in
magneto-optical traps, and high-quality compression of optical solitons \cite%
{anderson1994tunneling}.

As concerns the concept of \textquotedblleft shortcuts to adiabaticity"
(STA), it has recently drawn much interest to speed up slow adiabatic
processes, while suppressing the excitation or heating, with important
applications to atomic, molecular, optical and statistical physics, see
reviews \cite{torrontegui2013shortcuts,guery2019shortcuts,Kihwan}. In this
context, a series of works were devoted to frictionless
expansion/compression and cooling of atomic Bose-Einstein condensates (BECs)
in time-modulated harmonic-oscillator (HO) traps \cite%
{muga2009frictionless,chenprl104,schaff2010fast,schaff2011shortcut,del2011fast,schaff2011shortcuts}
, with extensions to cold-atom mixtures \cite{choi2011optimized},
Tonks-Girardeau (TG)\ \cite{adolfoTG,adolfoprx} and Fermi \cite%
{stringari,haibinpra} gases, and many-body systems \cite%
{guery2014nonequilibrium,rohringer2015non}. These results are not only
significant for the design of optimal quantum control \cite%
{rezek2009quantum,stefanatos2010frictionless}, but also have significant
implications for the studies of quantum speed limits, in the context of the
trade-off between time and energy cost under the constraint of the third law
in quantum thermodynamics \cite{hoffmann2011time,chen2010transient}. Other
systems, such as mechanical resonators \cite{li2011fast}, photonic lattices
\cite{stefanatos2014design}, bosonic Josephson junction \cite%
{julia2012fast,yuste2013shortcut,stefanatos2018maximizing,hatomura2018shortcuts}
, Brownian particles \cite{martinez2016engineered} and classical RC circuits
\cite{faure2019shortcut} have been extensively studied by using similar STA\
techniques for the swift transformation between two adiabatic or equilibrium
states.

Theoretically, the STA\ techniques, among which the most popular ones are
inverse engineering \cite{chenprl104}, counter-diabatic driving \cite%
{berry2009transitionless,adolfoprx,adolfocddriving} and fast-forward scaling
\cite{masuda2008fast,erikffward}, which were elaborated in different setups,
although they are mathematically equivalent \cite{erikffward,chen2011lewis}.
In the contexts of the inverse engineering, Lewis-Riesenfeld dynamical
invariant \cite{chenprl104}, or general scaling transformations \cite%
{castinprl,gritsev2010scaling}, various forms of the famous Ermakov equation
\cite{Ermakov0,Lewis,Ermakov1,Ermakov2,Ermakov3,Ermakov4} were derived for
designing shortcuts to adiabatic expansions of non-interacting thermal gases
and BEC in the Thomas-Fermi (TF)\ regime \cite%
{muga2009frictionless,schaff2011shortcut,del2011fast,schaff2011shortcuts}.
Specifically, when it comes to the shortcuts for BECs, the TF regime \cite%
{muga2009frictionless,schaff2011shortcuts} or time-dependent nonlinear
coupling \cite{muga2009frictionless} lead to (modified) Ermakov equations,
starting from the Gross-Pitaevskii (GP) equation, which is the commonly
adopted dynamical model of BEC in the mean-field theory \cite{Pitaevskii}.

In this work, inspired by approaches based on the variational approximation
(VA) \cite{variationalprl,juangoprl}, similar to those developed in
nonlinear optics \cite{anderson1994tunneling,malomed2002variational}, we
derive a generalized Ermakov equation, including a term induced by the
self-interaction term in the GP equation. The objective is to further
elaborate shortcuts for the adiabatic expansion/compression in BEC. This
allows us to manipulate nonlinear dynamics of BEC\ solitons by means of the
Feshbach resonance \cite{jingSciRep,jingnpj} and many-body dynamics in
power-law potentials \cite{xu2019quantum}. In particular, we exploit the VA
to design shortcuts to adiabaticity for the decompression of BEC in HO
traps. Exact solutions to the generalized Ermakov equation, including bang
and bang-bang control scenarios, are analytically obtained and used to
highlight the effect of inter-atomic interactions on the minimal time and
stability of the BEC manipulations. The results for the time-optimal driving
are different from those previously obtained for single atoms \cite%
{chen2010transient,stefanatos2010frictionless,rezek2009quantum} and BEC in
the TF limit \cite{optimalTF}, where negligible or very strong interactions
are assumed.

\section{The model, Hamiltonian, and variational approach}

We begin with the effective one-dimensional (1D) GP equation, modeling the
mean-field dynamics of the cigar-shaped BEC \cite{salasnichpra}:
\begin{equation}
i\frac{\partial \psi }{\partial t}=-\frac{1}{2}\frac{\partial ^{2}\psi }{
\partial x^{2}}+\frac{1}{2}\omega ^{2}(t)x^{2}\psi +gN|\psi |^{2}\psi ,
\label{GP-equation}
\end{equation}
where $\omega (t)$ is time-dependent trapping frequency, $g$ is the
nonlinearity coefficient representing atom-atom interaction, and $N$ is the
total number of atoms. The scaled variables are related to their
counterparts measured in physical units (with tildes) as per $t=\omega _{0}
\tilde{t}$, $\omega (t)=\tilde{\omega}(t)/\omega _{0}$, $x=\tilde{x}/\sigma
_{0}$, $g=\tilde{g}/\hbar \omega _{0}\sigma _{0}$, where $m$ is atomic mass,
$\omega _{0}$ is the initial longitudinal trapping frequency, $\sigma _{0}=
\sqrt{\hbar /m\omega _{0}}$ is the corresponding cloud size, $\tilde{g}
=2\hbar a_{s}\omega _{\perp }$ with scattering length $a_{s}$ and
the trapping frequency of the transverse potential, $\omega _{\perp }$,
which provides for reduction of the underlying three-dimensional GP equation
to the 1D form (\ref{GP-equation}), provided that  $\omega _{\perp }$ is much
larger than the one acting in the axial direction. Further, if the axial HO potential
is time-dependent, the use of the 1D equation (\ref{GP-equation}) is fully
justified if the respective frequencies of the time dependence are much smaller than
$\omega _{\perp }$.

In order to apply the VA \cite{variationalprl,juangoprl}, we start with the
Lagrangian density of Eq. (\ref{GP-equation}),
\begin{eqnarray}
\mathscr{L} &=&\frac{i}{2}\left( \psi \frac{\partial \psi ^{\ast }}{\partial
t}-\psi ^{\ast }\frac{\partial \psi }{\partial t}\right) -\frac{1}{2}%
\left\vert \frac{\partial \psi }{\partial x}\right\vert ^{2}  \notag \\
&&-\frac{1}{2}\omega ^{2}(t)x^{2}|\psi |^{2}-\frac{1}{2}gN|\psi |^{4}.
\label{Lagrangian density}
\end{eqnarray}%
Plugging the usual time-dependent Gaussian ansatz,
\begin{equation}
\psi (x,t)=A(t)\exp \left[ -\frac{x^{2}}{2a^{2}(t)}+ib(t)x^{2}\right] ,
\label{wavefunction}
\end{equation}%
in Eq. (\ref{Lagrangian density}), we calculate the effective Lagrangian $%
L=\int_{-\infty }^{+\infty }\mathscr{L} [\psi]dx$. Here $a(t)$ and $b(t)\ $
represent the width and chirp of the wave function, and amplitude $%
A(t)=(1/\pi a^{2})^{1/4}$ is the amplitude of wave function, determined by
the normalization condition, $\int_{-\infty }^{+\infty }\left\vert \psi
(x)\right\vert ^{2}dx=1$. The variational procedure applied to the
Lagrangian makes it possible to eliminate the chirp, $b=-\dot{a}/2a$, the
resulting Euler-Lagrange equation for $a(t)$ taking the form of the
generalized Ermakov equation \cite{quinnpra}:
\begin{equation}
\ddot{a}+\omega ^{2}(t)a=\frac{1}{a^{3}}+\frac{gN}{\sqrt{2\pi }a^{2}},
\label{Ermakov-like equation}
\end{equation}%
which is tantamount to the Newton's equation of motion for a particle with
unit mass, $\ddot{a}=-dU(a)/da$ (with the overdot standing for the time
derivative), with the effective potential and the corresponding energy,
\begin{eqnarray}
U(a) &=&\frac{1}{2}{\omega }^{2}(t)a^{2}+\frac{1}{2a^{2}}+\frac{gN}{\sqrt{%
2\pi }a},  \label{effective potential} \\
\mathcal{E}(a) &=&\frac{\dot{a}^{2}}{2}+\frac{1}{2}{\omega }^{2}(t)a^{2}+%
\frac{1}{2a^{2}}+\frac{gN}{\sqrt{2\pi }a}.  \label{effective energy}
\end{eqnarray}

The presence of term $\sim gN$ in Eq. (\ref{Ermakov-like equation}) makes it
different from the original Ermakov equation,
\begin{equation}
\ddot{a}+\omega ^{2}(t)a=\frac{1}{a^{3}},  \label{Ermakov equation}
\end{equation}
derived from the Lewis-Reseifiend invariant \cite{chenprl104} or by means of
the scaling transform \cite{castinprl,gritsev2010scaling}, see also Appendix \ref{appendix}.

\section{Shortcuts to adiabaticity}

In this section, we aim to construct STA protocols of time-dependent
trapping by selecting an appropriate time-dependent frequency in Eq. (\ref%
{Ermakov-like equation}), to guarantee a fast transform from the ground
state at time $t=0$ to another ground state at a fixed final time, $t=t_{f}$
, avoiding additional (unwanted) excitations. The initial value is taken as $%
\omega (0)=1$, and the final one is defined as $\omega (t_{f})=1/\gamma ^{2}$
, i.e., $\gamma \equiv \sqrt{\omega _{0}/\omega _{f}}$ may be considered an
appropriate control parameter. To guarantee that the initial and final
states are stationary ones, one has to impose the following boundary
conditions:
\begin{eqnarray}
a(0) &=&a_{\mathrm{i}},~~a(t_{f})=a_{\mathrm{f}},
\label{Bounday condition-1} \\
\dot{a}(0) &=&\dot{a}(t_{f})=0,  \label{Bounday condition-2} \\
\ddot{a}(0) &=&\ddot{a}(t_{f})=0,  \label{Bounday condition-3}
\end{eqnarray}
where $a_{\mathrm{i}}$ and $a_{\mathrm{f}}$ are unique positive real
solutions of equations
\begin{eqnarray}
a_{\mathrm{i}}^{4}-\frac{gN}{\sqrt{2\pi }}a_{\mathrm{i}} &=&1,  \label{ai} \\
\frac{a_{\mathrm{f}}^{4}}{\gamma ^{4}}-\frac{gN}{\sqrt{2\pi }}a_{\mathrm{f}
}&=&1,  \label{af}
\end{eqnarray}
which follow from the generalized Ermakov equation (\ref{Ermakov-like
equation}) with $\ddot{a}(0)=\ddot{a}(t_{f})=0$. Clearly, $a_{\mathrm{i}}=1$
and $a_{\mathrm{f}}=\gamma $ in the limit $gN\rightarrow 0$. Therefore, by
analogy to the perturbative Kepler problem, the boundary conditions defined
by Eqs. (\ref{Bounday condition-1})-(\ref{Bounday condition-3}) imply
minimization of the effective potential $U(a)$ (\ref{effective potential}),
as well as of the energy given by Eq. (\ref{effective energy}) without the
kinetic-energy term.

\subsection{Inverse engineering}

Here we address an example of atomic cooling by decompressing from initial
frequency $\omega _{0}=250\times 2\pi $ Hz to the final one $\omega
_{f}=2.5\times 2\pi $ Hz. In the linear limit, $gN\rightarrow 0$, the values
are $a_{\mathrm{i}}=1$ and $a_{\mathrm{f}}=\gamma =10$. However, due to the
atom-atom interaction, the initial and final sizes of the BEC cloud are
slightly different, $a_{\mathrm{i}}=1.001$ and $a_{\mathrm{f}}=10.099$
, as calculated numerically from Eqs. (\ref{ai}) and (\ref{af}) with the
nonlinearity strength $gN$. The boundary conditions being fixed, trajectory
of $a(t)$ may be approximated by the simplest polynomial ansatz \cite%
{chenprl104},
\begin{equation}
a(t)=a_{\mathrm{i}}-6(a_{\mathrm{i}}-a_{\mathrm{f}})s^{5}+15(a_{\mathrm{i}
}-a_{\mathrm{f}})s^{4}-10(a_{\mathrm{i}}-a_{\mathrm{f}})s^{3},
\label{ansatz}
\end{equation}
with $s=t/t_{f}$. As a consequence, smooth function $\omega (t)$ may be
\textit{inversely determined} by Eq. (\ref{Ermakov-like equation}). If an
imaginary trap frequency is dealt with, which corresponds to a parabolic
repeller, instead of the HO trap, in Eq. (\ref{GP-equation}), $t_{f}$ may be
formally made arbitrarily short. However, physical constraints always exist
in practice, see the discussion below. Generally, the use of the 
switch between the trapping and expulsive potentials extends possibilities 
for the design of control schemes with diverse functionalities.

\begin{figure}[t]
\scalebox{0.35}[0.35]{\includegraphics{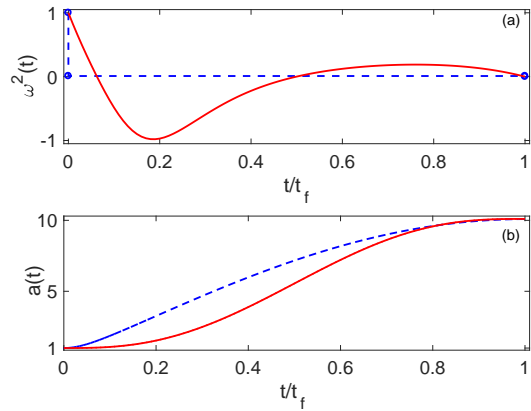}}
\caption{The time dependence of the designed trap frequency, $\protect\omega %
(t)$ (a), and width of the wave packet (b). Red solid and blue dashed lines
correspond to the inverse engineering and bang control, respectively. The
parameters are $\protect\omega (0)=1$, $\protect\omega (t_{f})=1/\protect%
\gamma ^{2}$, $gN=0.01$. Note that $t_{f}=5.45$ for the inverse engineering
is different $t_{f}=15.83$ for 2-jump control.}
\label{fig1}
\end{figure}

Here we chose $t_{f}=5.45$, such that the absolute value of the real
frequency is bounded by $\omega _{0}$ ($\omega _{0}<\omega _{f}$). Figure %
\ref{fig1} illustrates the respective time-varying trap frequency and
evolution of the width, as produced by the inverse-engineering method, where
the initial and final trap frequencies are $\omega (0)=1$ and $\omega
(t_{f})=1/\gamma ^{2}$ , and $gN=0.01$. Later, we will compare such smooth
trajectories with results produced by the so-called bang and bang-bang
control, which is relevant for the time-optimal solution, implemented by
means of piecewise-constant (discontinuous) intermediate trap frequencies.

\subsection{The two-jump control}

In the limit of $gN\rightarrow 0$, a simple exact solution of the Ermakov
equation can be constructed, that reproduces the shortcut with just one
intermediate frequency \cite{chen2010transient}, similar to the scenario for
the compression of solitons in nonlinear fibers, by passing the soliton from
a fiber segment with a large dispersion coefficient to a segment with is a
smaller one. This scenario was theoretically elaborated in Ref. \cite%
{anderson1994tunneling} and experimentally realized in \cite{experiment}.
Motivated by this, we assume that, at $t=0$, the trap frequency suddenly
changes from $\omega (0)$ to some constant intermediate value $\omega _{c}$,
to achieve an alternative shortcut.

In the linear limit, $gN\rightarrow 0$, the trap frequency remains equal to $%
\omega _{c}$, from $t=0$ to
\begin{equation}
t_{f}=\pi /\left( 2\omega _{c}\right) ,  \label{T}
\end{equation}
and at moment $t=t_{f}$ the frequency instantaneously changes from $\omega
_{c}$ to the final value, $\omega (t_{f})$. The exact solution of the
Ermakov equation (\ref{Ermakov equation}) with constant $\omega _{c}$ is
well known:
\begin{equation}
a^{2}(t)=\frac{1}{2}\left[ \left( A+C\right) +\left( A-C\right) \cos \left(
2\omega _{c}t\right) \right] ,  \label{solution}
\end{equation}
with constants $A$ and $C$ subject to constraint $AC=1/\omega _{c}^{2}$. To
secure the transformation of a stationary state taken at $t=0$ into another
stationary one at $t=t_{f}$, it is necessary to impose the above-mentioned
conditions, $a(0)=1$, $a(t_{f})=\gamma $, and $\dot{a}(0)=\dot{a}(t_{f})=0$.
After a straightforward algebra, the combination of such conditions and Eq.
( \ref{solution}) yields a simple solution:
\begin{equation}
a(t)=\sqrt{1+\frac{1-\omega _{c}^{2}}{\omega _{c}^{2}}\sinh ^{2}(\omega
_{c}t)},  \label{result}
\end{equation}
with
\begin{equation}  \label{bang}
\omega _{c}=\sqrt{\omega (0)\omega (t_{f})}=1/\gamma .
\end{equation}
Thus, Eqs. (\ref{T}) and (\ref{result}) provide a simple exact solution for
the shortcut if the nonlinearity is negligible. In particular, the necessary
intermediate trapping frequency $\omega _{c}$ is given as the geometric mean
of the initial and final frequencies.

The solution can be generalized for full equation (\ref{Ermakov-like
equation}), although in a less explicit form. The shortcut scenario implies
that the initial and final values (\ref{Bounday condition-1}), subject to
boundary conditions (\ref{Bounday condition-2}), are coupled by the motion
in potential (\ref{effective potential}). The energy conservation in this
mechanical (perturbative Kepler) problem implies $U(a_{\mathrm{i}})=U(a_{
\mathrm{f}})$, or, in an explicit form,
\begin{equation}
\omega _{c}^{2}=\frac{1}{a_{\mathrm{i}}^{2}a_{\mathrm{f}}^{2}}+\sqrt{\frac{2
}{\pi }}\frac{gN}{a_{\mathrm{i}}a_{\mathrm{f}}(a_{\mathrm{i}}+a_{\mathrm{f}%
}) }.  \label{tilde}
\end{equation}
In this case, a simple expression for $t_{f}$ is not available, but it can
be written in the form of an integral:
\begin{equation}
t_{f}=\int_{a_{\mathrm{i}}}^{a_{\mathrm{f}}}\frac{da}{\sqrt{2\left[ U(a_{
\mathrm{in}})-U(a)\right] }},  \label{bangt}
\end{equation}
where $a_{\mathrm{i}}$ and $a_{\mathrm{f}}$ are given by Eq. (\ref{Bounday
condition-1}). Thus, the trap frequency $\omega (t)$ and trajectory for $%
a(t) $ can be obtained in a numerical form from Eq. (\ref{Ermakov-like
equation}) with boundary conditions, see Fig. \ref{fig1}, where $\omega
_{c}=0.0993$ and $t_{f}=15.83$ are obtained for the chosen parameters, $%
\omega (0)=1$, $\omega (t_{f})=1/\gamma ^{2}$, and $gN=0.01$. Most
importantly, the designed trajectory $a(t)$ satisfies the boundary
conditions (\ref{Bounday condition-1}) and (\ref{Bounday condition-2}),
which guarantees the realization of the STA protocol and secures the
stability at $t>t_{f}$. However, since the boundary condition (\ref{Bounday
condition-3}), for the second derivative of $a(t)$, is not fulfilled, one
has to design the trap frequency suddenly change. Namely, the trap frequency
has to ``jump'' from initial value $\omega_{0}$ to intermediate one $%
\omega_c $ at $t=0$, and ``jump'' back to final one $\omega_{f}$ at $t=t_{f}$%
.

\subsection{The three-jump bang-bang control}

Next, we address the generalized Ermakov equation (\ref{Ermakov-like
equation}) and discuss the time-minimization optimal-control problem with a
constrained trap frequency, that is, $|\omega (t)|^{2}\leq \delta $. To
follow the usual conventions adopted in the optimal control theory, we set
new notation,
\begin{equation}
x_{1}\equiv a,~~ x_{2}\equiv \dot{a},  \label{xx}
\end{equation}
$u(t)\equiv \omega ^{2}(t)$, and rewrite Eq. (\ref{Ermakov-like equation})
as a system of the first-order differential equations:
\begin{eqnarray}
\dot{x}_{1} &=&x_{2},  \label{two-diffs} \\
\dot{x}_{2} &=&-ux_{1}+\frac{1}{x_{1}^{3}}+\frac{gN}{\sqrt{2\pi }}\frac{1}{
x_{1}^{2}},
\end{eqnarray}
where $x_{1}$, $x_{2}$ are the components of a \textquotedblleft state
vector" $\mathbf{x}$, and squared trap frequency $u(t)$ is considered as a
(scalar) control function. The form of the theoretical time-optimal solution
can be found using the Pontryagin's maximum principle, which provides
necessary conditions for the optimality. Similar to the approach used in
Refs. \cite{stefanatos2010frictionless,optimalTF}, determining the optimal
frequency profile reduces to finding $u(t)$ subject to the bound $|u(t)|\leq
\delta $ with $u(0)=1$ and $u(t_{f})=1/\gamma ^{2}$, such that the above
system starts with initial conditions $(x_{1}(0),x_{2}(0))=(a_{\mathrm{i}%
},0) $, and reaches the final point $(x_{1}(t_{f})$, $x_{2}(t_{f}))=(a_{%
\mathrm{f} },0)$ in minimal time $t_{f}$. The boundary conditions for $x_{1}$
and $x_{2} $ may be equivalently considered as those for $a$ and $\dot{a}$,
see Eqs. ( \ref{Bounday condition-1}) and (\ref{Bounday condition-2}). The
boundary conditions for $u(t)$ are equivalent to those for $\omega (t)$ and,
through Eq. (\ref{Ermakov-like equation}) or Eqs. (\ref{ai}) and (\ref{af}),
equivalent to those for $\ddot{a}$, hence there are, totally, six boundary
conditions, as in Eqs. (\ref{Bounday condition-1}-\ref{Bounday condition-3}).

To find the minimal time $t_{f}$, we define the cost function,
\begin{equation}
J=\int_{0}^{t_{f}}dt=t_{f}.
\end{equation}
The control Hamiltonian $H_{c}(\mathbf{p},\mathbf{x},u)$ is
\begin{equation}
H_{c}(\mathbf{p},\mathbf{x},u)=p_{0}+p_{1}x_{2}-p_{2}x_{1}u+\frac{p_{2}}{
x_{1}^{3}}+\frac{gN}{\sqrt{2\pi }}\frac{p_{2}}{x_{1}^{2}},
\end{equation}
where vector $\mathbf{p}=(p_{0},p_{1},p_{2})$ is composed of non-zero and
continuous Lagrange multipliers, $p_{0}<0$ may be chosen for convenience, as
it amounts to multiplying the cost function by a constant, and $p_{1,2}$
obey the Hamilton's equations: $\dot{\mathbf{x}}=\partial H_{c}/\partial
\mathbf{x}$ and $\dot{\mathbf{p}}=-\partial H_{c}/\partial \mathbf{x}$. For
almost all $0\leq t\leq t_{f}$, function $H_{c}(\mathbf{p},\mathbf{x},u)$
attains its maximum at $u=u(t)$, and $H_{c}(\mathbf{p},\mathbf{x},u)$ is a
constant. Making use of the Hamiltonian's equation, we arrive at the
following explicit expressions:
\begin{eqnarray}
\dot{p}_{1} &=&p_{2}\left( u+\frac{3}{x_{1}^{4}}+\frac{2gN}{\sqrt{2\pi }}
\frac{1}{x_{1}^{3}}\right) , \\
\dot{p}_{2} &=&-p_{1}.
\end{eqnarray}
It is clear that the control Hamiltonian $H_{c}(\mathbf{p},\mathbf{x},u)$ is
a linear function of variable $u(t)$. Therefore, the maximization of $H_{c}(
\mathbf{p},\mathbf{x},u)$ is determined by the sign of term $-up_{2}x_{1}$,
which, in turn, is determined by the sign of $p_{2}$, as $a(t)$ is always
positive, i.e., $x_{1}>0$ and $p_{2}\not=0$. Here $p_{2}=0$ does not provide
singular control, and only takes place at specific moments (switching
times). As a consequence, we arrive at the scheme of the \textquotedblleft
bang-bang" control, defined by the following form:
\begin{equation}
u(t)=
\begin{cases}
-\delta ,\qquad p_{2}>0, \\
~\delta ,~\qquad p_{2}<0,%
\end{cases}%
\end{equation}
which implies that the controller switches from one boundary value to the
other at the switching times. When $u$ is constant and Eq. (\ref%
{Ermakov-like equation}) holds, then it can be derived that
\begin{equation}
x_{2}^{2}+ux_{1}^{2}+\frac{1}{x_{1}^{2}}+\frac{2gN}{\sqrt{2\pi }x_{1}}=c,
\label{equation of x1 x2}
\end{equation}
where $c$ is an integration constant. Moreover, we see from Eq. (\ref%
{effective energy}) that trajectories with constant $u$ correspond to
constant energy $\mathcal{E}(a)=c/2$.

\begin{figure}[tbp]
\includegraphics[width=8cm]{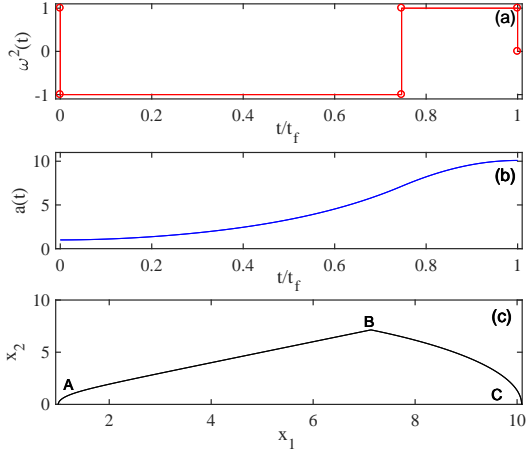}
\caption{Controller $u$ for \textquotedblleft bang-bang" scheme with one
intermediate switch (a), the width $a$ of the wave packet (b), and the
corresponding trajectory (c). Parameters: $\protect\delta =1$ and others are
the same as those in Fig. \protect\ref{fig1}.}
\label{fig2}
\end{figure}

By choosing the simplest but feasible \textquotedblleft bang-bang" control
with only one intermediate switching at $t=t_{1}$, we introduce the
three-jumps form,
\begin{equation}
u(t)=
\begin{cases}
1\qquad ~~~~~t=0, \\
-\delta ,\qquad ~~0<t<t_{1}, \\
\delta ,~~\qquad ~~t_{1}<t<t_{2}, \\
1/\gamma ^{2}\qquad ~t=t_{f}=t_{1}+t_{2},%
\end{cases}
\label{control sequence u(t)}
\end{equation}
as shown in Fig.~\ref{fig2}(a), where we take $\delta =1$ as an example, and
other parameters are the same as in Fig. \ref{fig1}. Obviously, the
discontinuities at the time edges are not implied by the maximum principle,
but are determined by the initial and final conditions imposed on control $u$
. This guarantees the creation of the STA protocol, but requires the sudden
change of the trap frequencies. Therefore, we can find a lower bound on the
minimum time, achieved only with instantaneous jumps of the control at the
initial and final times.

\begin{figure}[tbp]
\includegraphics[scale=0.35]{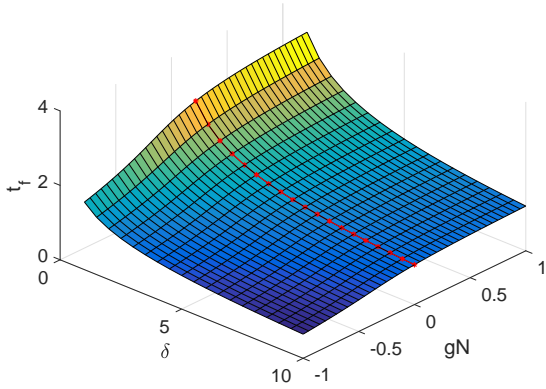}
\caption{The dependence of minimal time $t_{f}$ on bound $\protect\delta $
of the trap frequency and nonlinearity $gN$, other parameters being the same
as in Fig. \protect\ref{fig1}. The red dashed line indicates the minimal
time predicted by the optimal-control theory for $gN=0$.}
\label{fig3}
\end{figure}

Next, we aim to calculate the necessary time for the transfer from the
initial point, $A(a_{\mathrm{i}},0)$, to the final one, $C(a_{\mathrm{f}},0)$
, as shown in Fig. \ref{fig2}(c), where $B(x_{1}^{B},x_{2}^{B})$ is the
intermediate point at the switching instant, $t=t_{1}$. With the control
function taken as per Eq. (\ref{control sequence u(t)}), and boundary
conditions (\ref{Bounday condition-1}-\ref{Bounday condition-3}), we obtain
segment AB:
\begin{equation}
\dot{x}_{1}^{2}-\delta x_{1}^{2}+\frac{1}{x_{1}^{2}}+\frac{2gN}{\sqrt{2\pi }
x_{1}}=c_{1},  \label{first segment}
\end{equation}
with $c_{1}\equiv -\delta {a_{\mathrm{i}}}^{2}+1/a_{\mathrm{i}}^{2}+2gN/
\sqrt{2\pi }a_{\mathrm{i}}$, and segment BC:
\begin{equation}
\dot{x}_{1}^{2}+\delta x_{1}^{2}+\frac{1}{x_{1}^{2}}+\frac{2gN}{\sqrt{2\pi }
x_{1}}=c_{2},  \label{second segment}
\end{equation}
with $c_{2}\equiv \delta {a_{\mathrm{f}}}^{2}+1/a_{\mathrm{f}}^{2}+2gN/\sqrt{
2\pi }a_{\mathrm{f}}$. By using Eqs. (\ref{first segment}) and (\ref{second
segment}), the continuity condition at $t=t_{1}$ can be resolved for $%
x_{1}^{B}$ as follows:
\begin{equation}
x_{1}^{B}=\sqrt{\frac{1}{2}(a_{\mathrm{f}}^{2}+a_{\mathrm{i}}^{2})+\frac{%
(a_{ \mathrm{i}}^{2}-a_{\mathrm{f}}^{2})}{2\delta a_{\mathrm{f}}^{2}a_{%
\mathrm{i} }^{2}}+\frac{gN(a_{\mathrm{i}}-a_{\mathrm{f}})}{\sqrt{2\pi }%
\delta a_{ \mathrm{i}}a_{\mathrm{f}}}}.  \label{x1B}
\end{equation}
Once the intermediate point $x_{1}^{B}$ at switching time $t=t_{1}$ is
determined, we finally obtain
\begin{equation}
t_{f}=t_{1}+t_{2},  \label{tf for bangbang}
\end{equation}
where
\begin{eqnarray}
t_{1} &=&\int_{a_{i}}^{x_{B}}\frac{dx}{(\sqrt{\delta x^{2}-1/x^{2}+gN/\sqrt{
2\pi }x}+c_{1})},  \label{t1 for bangbang} \\
t_{2} &=&\int_{x_{B}}^{a_{f}}\frac{dx}{(\sqrt{-\delta x^{2}-1/x^{2}+gN/\sqrt{
2\pi }x+c_{2}})},  \label{t2 for bangbang}
\end{eqnarray}
Figures \ref{fig2}(b) and (c) illustrate the evolution of the soliton's
width and trajectory in phase space $(x_{1},x_{2})$, corresponding to
controller $u(t)$ of the \textquotedblleft bang-bang" type, where the
parameters are $\omega (0)=1$, $\omega (t_{f})=1/\gamma ^{2}$, $gN=0.01$ and
$\delta =1$. In this manner, we can find the minimal time for atomic
cooling, $t_{f}=3.097$, which is slightly larger than minimal time $%
t_{f}=3.088$ \cite{stefanatos2010frictionless,hoffmann2011time}, obtained
when in the linear limit, $gN=0$. In the opposite TF limit, the minimal time
$t_{f}=3.809$ is still large, see a detailed calculation in Appendix \ref{appendix}. Of
course, it may be possible to analyze the control strategy for schemes with
additional intermediate switchings, to predict shorter time for desired
transfer.

Figure \ref{fig3} shows the dependence of minimal time $t_{f}$ for the
\textquotedblleft bang-bang\textquotedblright\ control on the trap-frequency
bound $\delta $ and nonlinearity strength $gN$. On the one hand, the minimal
time approaches zero if there is no bound, i.e., $\delta \rightarrow \infty $
. On the other hand, minimal time $t_{f}$ is essentially affected by the
nonlinearity. The minimal time for the expansion of the condensate trapped
in the time-varying potential decreases with the increase of the
nonlinearity strength (through the Feshbach resonance). For instance, the
minimal time $t_{f}=3.079$ for $gN=-0.01$ is somewhat smaller than $%
t_{f}=3.088$ for $gN=0$, as indicated by the pointed line in Fig. \ref{fig3}
. Thus, the self-attractive (repulsive) nonlinearity accelerates the
expansion (compression) of the condensate. Furthermore, when $gN$ takes
larger values, BEC enters the TG regime. In this case, the scaling
transformation leads to the ordinary Ermakov equation, the minimal time
taking the above-mentioned value $t_{f}=3.088$, as the dynamics of the TG
gas can be reduced to a single-particle evolution, by dint of the Bose-Fermi
mapping. However, the system's fidelity and stability become quite different
when $gN$ changes from positive to negative values, as shown below.

\section{Discussion}

\subsection{Stability}

\begin{figure}[]
	\scalebox{0.36}[0.33]{\includegraphics{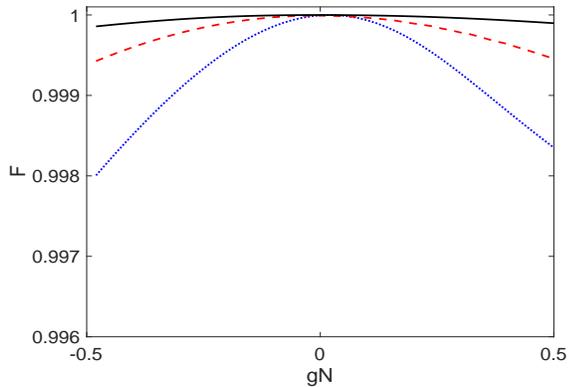}}
	\caption{The dependence of the fidelity, defined as per Eq. (\protect\ref{F}%
		) on nonlinearity strength $gN$, with the initial and final wave functions
		calculated by means of the imaginary-time evolution method. The state
		evolution, produced by different STA protocols, including inverse
		engineering (black solid), two-jump bang (blue dotted), and three-jump
		bang-bang (red dashed) schemes, is simulated with the help of the split-step
		method. Parameters are the same as those in Figs. \protect\ref{fig1} and
		\protect\ref{fig2}.}
	\label{fig4}
\end{figure}

\begin{figure}[]
	\scalebox{0.38}[0.38]{\includegraphics{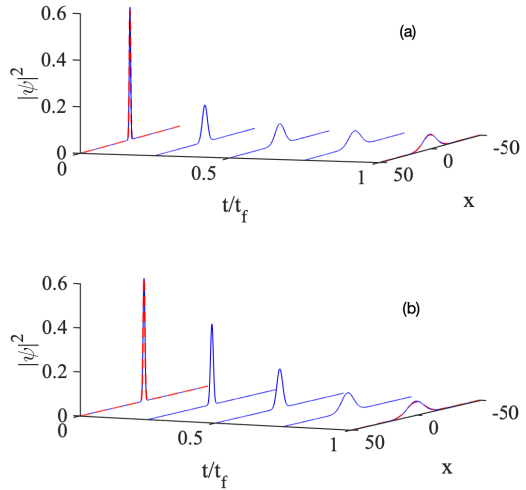}}
	\caption{The evolution of the particle density (squared 
		absolute value of the wave function) governed by the GP equation (%
		\protect\ref{GP-equation}), with trap frequency $\protect\omega (t)$
		modulated in time by (a) the two-jump (\protect\ref{bang}) and (b) the
		three-jump bang-bang (\protect\ref{control sequence u(t)}) controls.
		Red dashed lines represent the initial and final
		particle densities calculated as per the variational approximation.
		They are virtually indistinguishable from the counterparts produced 
		by the numerical solution of the GP equation.   
		Parameters are the same as those in Fig. \protect\ref{fig2}.}
	\label{fig5}
\end{figure}

Here we aim to explore stability of different STA protocols, designed on the
basis of the inverse engineering, two-jump and three-jump bang-bang schemes,
against the variation of the nonlinearity strength $gN$. To this end, we
calculate the fidelity defined as
\begin{equation}
F=|\langle \tilde{\psi}_{f}(x)|\psi (x,t_{f})\rangle |^{2},  \label{F}
\end{equation}%
where wave function $|\tilde{\psi}_{f}\rangle $ is the final stationary
state. Here the imaginary-time evolution method is used for obtaining the
initial and final stationary states, and the state evolving along the
shortcut trajectory is numerically calculated by means of the split-step
method. As illustrated by Fig. \ref{fig4}, the smooth STA\ trajectory
designed by dint of inverse engineering demonstrates, in general, better
tolerance against the nonlinearity effects (which make the fidelity poorer),
as two or three-jump protocols require abrupt changes of the frequency to
satisfy the boundary conditions, which is a challenging condition. The STA
protocols are more stable for $gN>0$ than for the opposite sign, as the
Newtonian particle can easier escape from the effective potential well when
the nonlinearity is self-attractive. As a matter of fact, the nonlinearity
strongly affects the initial and final sizes of the BEC cloud (\ref{Bounday
condition-1}). Under the action of weak nonlinearity, the difference between
stationary states produced by the imaginary-time evolution method and the
assumed Gaussian wave packets with initial and final values of $a$ [see Eq.
( \ref{wavefunction})] is negligible. Furthermore, Fig. \ref{fig5} shows the
time evolution of the particle density, produced by the numerical solution of
the time-dependent GP equation (\ref{GP-equation}), and its counterpart 
predicted by the VA (dashed red curves). The figure corroborates
the validity of the VA based on Gaussian ansatz (\ref{ansatz}), while the
trap frequency changes abruptly at the switching points.

Apart from that, the implementation of our proposed protocols in realistic
BEC experiments requires careful considerations. First, one has to apply two
pinch coils to offset a purely magnetic Ioffe-Pritchard trap, thus producing
an expulsive quadratic potential, instead of the HO one \cite{salomon}. An
alternative way for achieving the same purpose is to combine a
time-dependent red-detuned optical dipole trap with an additional
blue-detuned antitrap \cite{chenprl104}. Second, sudden changes of on-off
controller entail the fast trap modulation, which might lead to unwanted
intrinsic excitation of the state under the consideration \cite{carrpra}. To
avoid this, a multiple shooting method should be used for smoothing the
bang-bang control scheme \cite{yongchengpra}. It is also important to
mention that our idealized model amounts to an effectively 1D
trap, produced by integrating the underlying three-dimensional GP equation
in the transverse directions, under the action of the confinement in the
transverse plane \cite{salasnichpra}. The use of magnetic and optical traps
allows one to independently control of the axial and transverse frequencies.
In fact, the radial-longitudinal coupling in the 3D setting sets a limit for
the time scale on which the 1D equation is valid. It may be
improved by increasing the waist of the trapping laser beam \cite{torronteguiPRA2012}. To be more precise, by taking into account
longitudinal anharmonic perturbations, the lower validity bound for the
ground-state decompression in the Gaussian trap is found to be $t_{f}\gg
3\hbar /8mw^{2}\omega _{f}^{2}$ \cite{xiaojingpra14,chenprl104}, with $m$
and $w$ being the mass of $\mbox{Rb}^{87}$ atoms and the laser-beam's waist.
Thus, after choosing the initial and finial frequencies, we obtain $t_{f}\gg
0.7$ for $w=50~\mathrm{\mu }$m and $t_{f}\gg 0.08$ for $w=150~\mathrm{\mu }$%
m, showing the validity of different STA protocols with high fidelity under
realistic conditions.

\subsection{Excitation energy}

The STA protocols support the frictionless cooling subject to the initial
and final boundary conditions. However, the process itself is not adiabatic
at all, thus excitation of the system on top of the stationary state may
lead to detrimental effects. To address this issue, we define the
time-average energy as
\begin{equation}
\bar{\mathcal{E}}=\frac{1}{t_{f}}\int_{0}^{t_{f}}\mathcal{E}(t)dt.
\label{time average E}
\end{equation}
Substituting Eq. (\ref{effective energy}), and integrating once with the use
of boundary conditions (\ref{Bounday condition-1})-(\ref{Bounday condition-3}
), we obtain
\begin{equation}
\bar{\mathcal{E}}=\frac{1}{t_{f}}\int_{0}^{t_{f}}\left( \dot{a}^{2}+\frac{1}{
a^{2}}+\frac{3gN}{2\sqrt{2\pi }a}\right) dt.
\end{equation}
The dependence of the so computed excitation energy on the nonlinearity
strength is presented in Fig. \ref{fig5}. In principle, such an energy price
of STA protocols is stipulated by the time-energy uncertainty, which implies
increase of the (time-averaged) energy for shorter times. It is seen that
the two-jump bang scheme produces smaller excitation energy, as the
corresponding operation time is larger. With the same frequency bound, the
operation time for the inverse-engineering scheme is larger than for the
time-optimal bang-bang one, which leads to a smaller excitation energy as
well. One can use another ansatz for the inverse engineering to minimize the
excitation energy, as discussed in work \cite{chen2010transient}. In
addition, we point out that, as the bang and bang-bang schemes require
sudden jumps to match the boundary conditions, the extra energy cost has to
be paid at the edges, to fully implement these schemes.

\begin{figure}[tbp]
\scalebox{0.36}[0.33]{\includegraphics{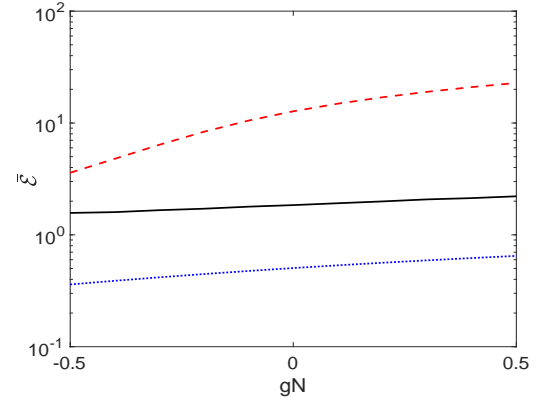}}
\caption{The dependence of time-average energy $\bar{\mathcal{E}}$, defined
by Eq.~(\protect\ref{time average E}), on the nonlinearity strength $gN$,
for the inverse-engineering (black solid), two-jump bang (blue dotted), and
three-jump bang-bang (red dashed) schemes. Parameters are the same as in
Figs. \protect\ref{fig1} and \protect\ref{fig2}.}
\label{fig6}
\end{figure}

\begin{figure}[tbp]
\scalebox{0.36}[0.33]{\includegraphics{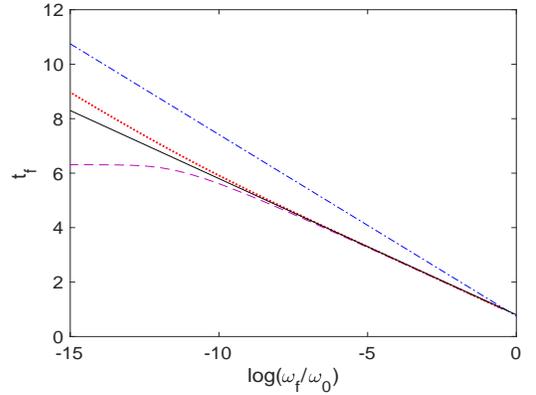}}
\caption{The minimum time for bang-bang scheme, as a function of $\log (
\protect\omega _{f}/\protect\omega _{0})$: the TF limit (the dashed-dotted
blue line), $gN=0$ (the solid black line), $gN=0.01$ (the dotted red line),
and $gN=-0.01$ (the dashed purple line), other parameters being the same as
in Fig. \protect\ref{fig2}. }
\label{fig7}
\end{figure}

Finally, we note that the STA approach for atom cooling has fundamental
implications for the third law of thermodynamics, with the atoms being the
medium in a quantum refrigerator. Figure \ref{fig7} quantifies the third law
in this case, i.e., the minimal time diverges when the final trap frequency $%
\omega _{f}$, proportional to temperature, approaches zero. In a more
general case, we see from Fig. \ref{fig7} that the unattainability principle
is quantified as
\begin{equation}
t_{f}\propto \log (a_{\mathrm{f}})+\frac{\pi }{4},
\end{equation}
where $a_{\mathrm{f}}$ is a positive real solution of Eq. (\ref{af}), and we
keep the first two terms of its Taylor's expansion around $gN=0$,
\begin{equation}
a_{\mathrm{f}}=\left( \frac{\omega _{0}}{\omega _{f}}\right) ^{1/2}+\frac{gN
}{4\sqrt{2\pi }}\left( \frac{\omega _{0}}{\omega _{f}}\right)
^{3/2}+O^{2}(gN).  \label{expansion}
\end{equation}
Interestingly, from Eq. (\ref{expansion}) we recover the scaling law, $%
t_{f}\propto (\omega _{0}/\omega _{f})^{1/2}$, when $gN=0$, leading to the
cooling rate, $R\propto T^{3/2}$, of the quantum refrigerator \cite%
{salamon2009maximum,hoffmann2011time,chen2010transient}. In the TF limit,
the second term in Eq. (\ref{expansion}) becomes dominant, yielding $%
t_{f}\propto (\omega _{0}/\omega _{f})^{3/2}$, with the corresponding
cooling rate $R\propto T^{2}$. Remarkably, the self-repulsive nonlinearity
provides a larger exponent, therefore the repulsive interaction, acting in
the course of the cooling process, may improve the cooling rate. When the
nonlinear Feshbach heat engine \cite{jingnpj} is considered, the attractive
self-interaction implies the shorter time, leading to the improvement of the
work. But in this case the stability is weaker, and the collapse of the wave
packet exists at $\omega _{f}\ll 1$.

\section{Conclusion}

To summarize, we have discussed the STA (shortcut to adiabaticity) for the
expansion of weakly interacting BEC loaded in the HO (harmonic-oscillator)
trap. The analysis is based on the use of the generalized Ermakov equation,
derived from the VA (variational approximation) applied to the effective 
1D Gorss-Pitaevskii equation. Exact solutions of the
generalized Ermakov equation are \textit{inverse engineered} to design
smooth or piecewise-constant intermediate time dependences of the trap
frequency, which realize the STA schemes. In particular, we focused on the
minimal transition time and the minimization of the excitation energy
produced by STA. To this end, the time-optimal solutions provided by the
bang-bang scheme with a smooth polynomial ansatz, and by the two-jump bang
scheme, have been compared. We conclude that the self-attractive
nonlinearity in BEC can help to shorten the minimal time. The latter result
which may have fundamental implications for the consideration of the quantum
speed limit and third law of thermodynamics in quantum systems.

Finally, we point out several pending issues to be addressed. The stability
with respect to the intrinsic BEC nonlinearity is reasonable, but the
validity of the VA derivation of the Ermakov equation is predicated upon the
accuracy of the Gaussian ansatz. Definitely, the bright soliton based on the
hyperbolic-tangent function may be another option. Another noteworthy point
is that the time-optimal solution has been obtained by means of the
\textquotedblleft bang-bang" scheme. It requires the sudden change of the
trap frequency, that may be difficult to implement physically. One can
further optimize the trajectory with more constraints imposed on the first,
or even second, derivatives of the time dependence of the trap frequency. An
alternative may be to use a smooth ansatz with polynomial and trigonometric
functions for constructing the time-optimal \textquotedblleft bang-bang"
scheme, as discussed in Ref. \cite{martikyan2019comparison}. All these
results may apply to analyzing the transport, splitting, and compression of
solitons \cite{jingSciRep,jingnpj}, and also to various anharmonic
potentials \cite{xu2019quantum,xiaojingpra14}, e.g. with quadratic terms.

\section*{Acknowledgments}

We acknowledge support from National Natural Science Foundation of China
(NSFC) (11474193), STCSM (2019SHZDZX01-ZX04, 18010500400 and 18ZR1415500),
Program for Eastern Scholar, Ram\'on y Cajal program of the Spanish MCIU
(RYC-2017-22482), QMiCS (820505) and OpenSuperQ (820363) of the EU Flagship
on Quantum Technologies, Spanish Government PGC2018- 095113-B-I00
(MCIU/AEI/FEDER, UE), Basque Government IT986-16, as well as the and EU FET
Open Grant Quromorphic. The work of BAM is supported, in part, by the Israel
Science Foundation, through grant No. 1286/17.

\section*{Data Availability Statement}

The data that support the findings of this study are available from the
corresponding author upon reasonable request.

\bibliographystyle{apsrev4-1}
\bibliography{chaos}

\appendix
\section{}
\label{appendix}

In the appendix, we first present an alternative way to derive the minimal
time in TF limit for the consistence and further comparison. We use a set of
equations for the density, $n$, and phase gradient, $\bigtriangledown \phi
\left( x,t\right) $, of the wave function, represented in the Madelung form,
\begin{equation}
\psi \left( x,t\right) =\sqrt{n\left( x,t\right) }e^{i\phi \left( x,t\right)
}.  \label{wave fun for n}
\end{equation}
Thus, the continuity equation derived from the GP equation (\ref{GP-equation}
) is
\begin{equation}
\frac{\partial n\left( x,t\right) }{\partial t}+\nabla \cdot \left( n\mathbf{%
\ v}\right) =0,  \label{continuity equation}
\end{equation}
where the velocity of the superflow is defined by
\begin{equation}
\mathbf{v}=\frac{1}{2i}\frac{\left( \psi ^{\ast }\nabla \psi -\psi \nabla
\psi ^{\ast }\right) }{|\psi |^{2}}\equiv \nabla \phi .  \label{define v}
\end{equation}
Next, we insert expression $(\ref{wave fun for n})$ in Eq. $(\ref%
{GP-equation})$, the real part of which yielding the Euler equation:
\begin{equation}
\frac{\partial \mathbf{v}}{\partial t}=-\bigtriangledown \left[ -\frac{1}{2}
\frac{\bigtriangledown ^{2}\sqrt{n}}{\sqrt{n}}+\frac{\mathbf{v}^{2}}{2}+
\frac{1}{2}\omega ^{2}\left( t\right) x^{2}+gn\right] .
\label{Euler equation by hydro}
\end{equation}
In the case of the strongly self-repulsive condensate, the TF approximation
\cite{Pitaevskii} allows one to omit the kinetic-energy term in Eq. $(\ref%
{Euler equation by hydro})$ (the first term of right-hand side), thus
neglecting the quantum pressure, the result being
\begin{equation}
\frac{\partial \mathbf{v}}{\partial t}+\frac{\partial }{\partial x}\left(
\frac{\mathbf{v}^{2}}{2}+\frac{1}{2}\omega ^{2}\left( t\right)
x^{2}+gn\right) =0  \label{Euler equation of TF}
\end{equation}
In the TF approximation, the initial equilibrium density distribution is $%
n_{0}(x,t)=(\mu -\omega ^{2}(0)x^{2}/2)/g$, according to the
time-independent GP equation with chemical potential $\mu $. The scaling
approach to the hydrodynamic equation is commonly used to study dynamical
properties of cold atomic system. It is based on ansatz of $%
n(x,t)=n_{0}[x/a(t)]/a(t)$, which satisfies the initial condition $%
n(x,0)=n_{0}(x)$ and $a(0)=1$. Then, one obtains the velocity field from the
continuity equation $(\ref{continuity equation})$:
\begin{equation}
\mathbf{v}=\frac{\dot{a}\left( t\right) }{a\left( t\right) }x
\label{scaling of v}
\end{equation}
Combining the TF limit and the scaling approach by inserting expression $( %
\ref{scaling of v})$ in Eq. $(\ref{Euler equation of TF})$ produces the
evolution equation for scaling factor $a(t)$:
\begin{equation}
\ddot{a}+\omega ^{2}\left( t\right) a=\frac{\omega _{0}^{2}}{a^{2}}.
\label{Ermakov for TF}
\end{equation}

Next, we need to transfer the initial ground state from trap frequency $%
\omega (0)=1$ at $t=0$ to the target state with $\omega (t_{f})=1/\gamma
^{2} $. And the same boundary conditions, Eqs. (\ref{Bounday condition-1}-%
\ref{Bounday condition-3}), where $a_{\mathrm{i}}=1$ and $a_{\mathrm{f}%
}=\gamma ^{4/3}$, are imposed to guarantee the realization of STA.
Then, in terms of notation (\ref{xx}), the dynamical equations are rewritten
as
\begin{eqnarray}
\dot{x}_{1} &=&x_{2}, \\
\dot{x}_{2} &=&-\tilde{u}x_{1}+\frac{1}{x_{1}^{2}},
\end{eqnarray}
where the control function $\tilde{u}(t)\equiv \omega ^{2}(t)/\omega
_{0}^{2} $ is used, subject to the bound $|\tilde{u}(t)|\leq \delta $, and
variables $x_{1,2}$ satisfies obey the constraint
\begin{equation}
x_{2}^{2}+\tilde{u}x_{1}^{2}+\frac{2}{x_{1}}=c,
\end{equation}
where $c$ is an integration constant. The time-optimal solution of the
\textquotedblleft bang-bang\textquotedblright\ type \cite{optimalTF} for $%
\tilde{u}(t)$ is built as the same as $u(t)$ in (\ref{control sequence u(t)}%
). 
Applying the calculations similar to those presented in Eqs. $(\ref{first
segment})$-$( \ref{tf for bangbang})$, the expressions for the duration of
these two segments reads
\begin{equation}
t_{1}=\int_{a_{\mathrm{i}}}^{x_{1}^{B}}\frac{dx_{1}}{\sqrt{\delta
x_{1}^{2}-2/x_{1}+c_{1}}},
\end{equation}
\begin{equation}
t_{2}=\int_{x_{1}^{B}}^{a_{\mathrm{f}}}\frac{dx_{1}}{\sqrt{-\delta
x_{1}^{2}-2/x_{1}+c_{2}}},  \label{t_2 for TF}
\end{equation}
with $c_{1}=2/a_{\mathrm{i}}-\delta a_{\mathrm{i}}^{2}$ and $c_{2}=\delta
a_{ \mathrm{f}}^{2}+2/a_{\mathrm{f}}$. The intermediate switching point $%
x_{1}^{B}$ is found as
\begin{equation}
x_{1}^{B}=\sqrt{\frac{a_{\mathrm{f}}^{2}+a_{\mathrm{i}}^{2}}{2}+\frac{a_{
\mathrm{i}}-a_{\mathrm{f}}}{\delta a_{\mathrm{i}}a_{\mathrm{f}}}}.
\end{equation}
The minimum time in the TF limit, $t_{f}$, is now $t_{f}=t_{1}+t_{2}$.

Finally, we briefly present the time-optimal scheme for the ordinary Ermakov
equation $(\ref{Ermakov equation})$. In this case, the boundary conditions,
Eqs. (\ref{Bounday condition-1}-\ref{Bounday condition-3}), hold with $a_{%
\mathrm{i}}=1$ and $a_{\mathrm{f}}=1/\gamma $. Following Ref. \cite%
{stefanatos2010frictionless}, the minimal time $t_{f}=t_{1}+t_{2}$ is
obtained, where the switching times $t_{1}$ and $t_{2}$ are
\begin{equation}
t_{1}=\frac{1}{\sqrt{\delta }}\sinh ^{-1}\sqrt{\frac{(\gamma ^{2}-1)(\gamma
^{2}\delta -1)}{2\gamma ^{2}(1+\delta )}},
\end{equation}
\begin{equation}
t_{2}=\frac{1}{\sqrt{\delta }}\sin ^{-1}\sqrt{\frac{(\gamma ^{2}-1)(\gamma
^{2}\delta +1)}{2(\gamma ^{4}\delta -1)}}.  \label{t_2 for Ermarkov}
\end{equation}
By setting $\delta =1$, we eventually obtain
\begin{eqnarray}
t_{1} &=&-\frac{1}{r2}\log \left( \frac{\omega _{0}}{\omega _{f}}\right) , \\
t_{2} &=&\pi /4,
\end{eqnarray}
as indicated in Fig. \ref{fig7}.

\end{document}